\documentclass[10pt, a4paper, onecolumn]{article}

\usepackage[letterpaper,top=2cm,bottom=2cm,left=3cm,right=3cm,marginparwidth=1.75cm]{geometry}

\usepackage[numbers]{natbib}%very important
\usepackage[english]{babel}
\usepackage[T1]{fontenc}
\usepackage{lmodern}
\usepackage{natbib}
\usepackage[utf8]{inputenc} % allow utf-8 input
\usepackage{nicefrac}       % compact symbols for 1/2, etc.
\usepackage{microtype}      % microtypography
\usepackage{bm}
\usepackage{url}

\PassOptionsToPackage{table}{xcolor}
\usepackage{graphicx}
\usepackage{multirow}
\usepackage{mathtools}
\usepackage[flushleft]{threeparttable} % http://ctan.org/pkg/threeparttable
\usepackage{booktabs}
\usepackage[table]{xcolor}  % If error use \PassOptionsToPackage{table}{xcolor} before \documentclass{...}
\usepackage{array}
\usepackage{balance}
\newcommand{\figref}[1]{Figure~\ref{#1}}

\newcommand{\tblref}[1]{Table~\ref{#1}}
\newcommand{\secref}[1]{Section~\ref{#1}}

\newcommand{\pz}{\phantom{0}}

\newcommand{\AP}[1]{\ensuremath{\text{AP}_{#1}}}
\newcommand{\mAP}[1]{\ensuremath{\text{mAP}_{#1}}}
\newcommand{\AR}[1]{\ensuremath{\text{AR}_{#1}}}
\newcommand{\mAR}[1]{\ensuremath{\text{mAR}_{#1}}}

\newcommand{\Description}[1]{}

\title{Multiperspective Teaching of Unknown Objects via Shared-gaze-based Multimodal Human-Robot Interaction}

\author{
	Daniel Weber%\thanks{Funded by the Deutsche Forschungsgemeinschaft (DFG, German Research Foundation) under Germany’s Excellence Strategy -- EXC number 2064/1 -- Project number 390727645.}
	\\ University of Tübingen \\ \texttt{daniel.weber@uni-tuebingen.de}
	\and
	Wolfgang Fuhl \\ University of Tübingen \\ \texttt{wolfgang.fuhl@uni-tuebingen.de}
	\and
	Enkelejda Kasneci \\ Technical University of Munich \\ \texttt{enkelejda.kasneci@tum.de}	
	\and
	Andreas Zell \\ University of Tübingen \\ \texttt{andreas.zell@uni-tuebingen.de}
}

\begin{document}

	\maketitle
	
	\begin{abstract}
		For successful deployment of robots in multifaceted situations, an understanding of the robot for its environment is indispensable.
With advancing performance of state-of-the-art object detectors, the capability of robots to detect objects within their interaction domain is also enhancing.
However, it binds the robot to a few trained classes and prevents it from adapting to unfamiliar surroundings beyond predefined scenarios. In such scenarios, humans could assist robots amidst the overwhelming number of interaction entities and impart the requisite expertise by acting as teachers.
We propose a novel pipeline that effectively harnesses human gaze and augmented reality in a human-robot collaboration context to teach a robot novel objects in its surrounding environment.
By intertwining gaze (to guide the robot's attention to an object of interest) with augmented reality (to convey the respective class information) we enable the robot to quickly acquire a significant amount of automatically labeled training data on its own.
Training in a transfer learning fashion, we demonstrate the robot's capability to detect recently learned objects and evaluate the influence of different machine learning models and learning procedures as well as the amount of training data involved. 
Our multimodal approach proves to be an efficient and natural way to teach the robot novel objects based on a few instances and 
allows it to detect classes for which no training dataset is available.
%even allows it to generalize to unseen objects in the given class.
In addition, we make our dataset publicly available to the research community, which consists of RGB and depth data, intrinsic and extrinsic camera parameters, along with regions of interest.

	\end{abstract}	
	
	\section{Introduction}
As technology progressed, more and more robots were developed for the industrial sector, and their fields of application became diversified.
Numerous industries, including automotive, electronics, rubber and plastics, cosmetics, pharmaceutical, and food and beverage, benefit from their superior precision, efficiency, working capacity and tolerance to arduous and hazardous environments~\cite{xiao2021robotics}.
In the immediate environment of humans, robots are also increasingly employed in the form of service assistants, for instance in supermarkets such as Walmart~\cite{bogue2019strong,li2022effect} or as tour guides in museums~\cite{thrun1999minerva, velentza2019human}.
This success has also been fueled by recent advances in machine learning, particularly in computer vision, which allows robots to understand their environment and detect objects and people within it.
However, a core assumption is almost always that a large amount of training data with high quality labels exist.
Many large car manufacturers or companies such as Google, Tesla, and Uber have therefore established their own image and video databases, in most cases by outsourcing to crowdsourcing platforms such as Amazon Mechanical Turk~\cite{schmidt2019crowdsourced}. % located in countries with cheap human labor
Regarding service robots operating in warehouses or office environments, there are often no publicly accessible datasets that are tailored to the respective environment, comprise all relevant object classes, and are fully labeled.
Consequently, state-of-the-art object detectors perform excellently given the existence of sufficient training data, but are limited to deployment in previously specified scenarios predefined by the training data \cite{weber2020distilling}.
As soon as an object is not included, it exceeds the capabilities of the object detector and pushes the robot to the limits of its possibilities.
In fact, the proportion of objects covered by data sets is vanishingly small compared to the quantity of objects existing in practice.
For example, ImageNet~\cite{deng2009imagenet}, one of the largest publicly available image datasets, contains just 1000 classes, while the number of classes existing in the real world obviously far exceeds this number.
This fact hinders the deployment of robots in unknown environments and the dynamic adaptation to unfamiliar conditions.

In this work, we aim to make mobile robots not only more capable in terms of the tasks they have to accomplish, but also alleviate data dependency, in the sense that we extend the robot's basic knowledge by building on an existing general understanding of objects and adding new classes.
More specifically, we
%address this problem by teaching
teach a robot novel, unknown objects to enable it to redetect said objects within the environment in which the learning process took place.
To this end, we employ fluent and intelligent human-robot interaction, at the intersection of research fields of computer vision, eye tracking, and augmented reality (AR).
By means of the latter, we realize a multimodal communication channel, using human gaze to direct attention to an object, and speech or gestures to convey the relevant class information to the robot.
Subsequently, the robot visually segments the object of interest and takes a series of images of the object from slightly different angles.
The data obtained in this user-friendly and convenient process is rich in information and encompasses extrinsic and intrinsic camera parameters, as well as regions of interest, in addition to RGB and depth images.
In conjunction with the class information provided by the human, the robot learns the respective object accordingly.

In summary, our main contributions are as follows:
\begin{enumerate}
    \item We propose a novel pipeline to teach a robot new, yet unknown objects.
    \item Towards this goal, we combine gaze and augmented reality in a human-robot interaction scenario to enable a feasible and swift acquisition of large amounts of labeled training data.
    \item We evaluate the learning process in detail with multiple models, different learning methods, and with varying amounts of data.
    % \item We make our versatile dataset publicly available to the research community under BLIND.
    \item We present Objects in Multiperspective Detail (OMD), a versatile dataset, and make it publicly available to the research community under \url{https://cloud.cs.uni-tuebingen.de/index.php/s/2oRPs2o3FZkdBHW}.
    \item We make our system with our entire code base publicly available to the research community under \url{https://github.com/dnlwbr/Multiperspective-Teaching}.
\end{enumerate}

	\section{Related Work}
Engaging multimodal human-robot interaction to teach a robot unknown objects requires significant multidisciplinary efforts, which we will discuss in this section. From
1) augmented reality in robotics, and
2) previous applications of eye tracking in the context of computer vision, to
3) unknown object detection, to
4) how robots learn and
5) collaborate with humans.

\subsection{Augmented Reality in Robotics}
With the increasing popularity of augmented reality devices, industrial applications and research also expanded~\cite{makhataeva2020augmented}.
Especially the combination of AR, %that enhances the real world with digital visual elements,
with robotic-assisted surgeries showed potential~\cite{qian2019review}, as the human remains in control via AR, but can take advantage of the precision and consistency of the robots~\cite{vadala2020robotic}.
In terms of robot control and path planning, \cite{krupke2018comparison} controlled an industrial robot arm in pick-and-place tasks using an AR device and \cite{quintero2018robot} designed an AR interface to plan, preview and execute the trajectory of a robot arm.
In multi-agent systems, AR has also been used for visual feedback~\cite{von2016robot} and remote control~\cite{reina2017ark} of robot swarms.
Enhancing the perception of the real world through AR has thus proven to be an appealing way to communicate with robots.

\subsection{Eye Tracking and Computer Vision}
Eye tracking has also become an important tool in both research and industry~\cite{krafka2016eye}.
For this reason, \cite{krafka2016eye} developed an eye tracking software that works on mobile devices such as mobile phones or tablets and does not require any sensors other than a camera.
The authors of \cite{pfeiffer2014eyesee3d}, analyzed the mobile 3D eye tracking data using computer vision (and augmented reality). This involved tracking 3D markers and aligning them with virtual proxies.
% "Our approach combines geometric modelling with inexpensive 3D marker tracking to align virtual proxies with the real-world objects."
In~\cite{toyama2012gaze}, the class of objects was identified by matching the features of known objects with features in the neighborhood of human fixations.
However, this required the object to appear in the database previously created specifically for this purpose, and it was also not possible to make statements about the position of the objects.
%By using a learning model and multiple fixation points, \cite{xiao2018salient} reduced the number of superpixels for salient object detection.
Nevertheless, eye tracking data can help to improve the performance of segmentation algorithms~\cite{mishra2009active}.
Thus, in~\cite{papadopoulos2014training}, fixations were used to train a model to annotate object locations, and in \cite{weber2022exploiting} gaze was incorporated into the segmentation process of a point cloud.
Here in turn, in both cases it was not possible to make a statement about the object classes.

\subsection{Unknown Object Detection}
The problem of unknown object detection was also addressed using eye tracking in~\cite{weber2020distilling}.
In this work, the number of candidate bounding boxes of a region proposal method was significantly reduced, but a classification was not possible.
Using heat maps instead of scene frames, \cite{weber2022gaze} categorized video segments based on whether a person was looking at an object and determined the parameters of the associated bounding box.
The detected objects were all unknown, but again the classes were not determined.
The authors of \cite{kabir2022unknown} addressed the problem of unknown object detection using a one-class support vector machine. 
Since the learning process was incremental, multiple robots were involved, connected to each other via a cloud-based station where all the processing took place.
In this approach, unknown objects were only identified as unknown without adding the class information to the learning process.
The purpose was to filter the unknown objects from the classified objects and forward only known objects in order to avoid sending incorrect information to the robots.
In another recent work, \cite{li2022uncertainty} proposed to exploit additional predictions of semantic segmentation models and quantify the uncertainty of the proposed segmentations.
Again, the classification task was binary and only the categories known and unknown were determined without eventually learning the objects.

\subsection{Teaching of Robots and Machines}
Robots are employed in a wide range of applications, especially in industry.
These include teaching assembly tasks \cite{wan2017teaching}, where the robot learns from human demonstration:
First the robot observes the human, then it imitates the human's movements.
A different way of learning through user interaction is proposed in \cite{gemignani2015teaching}.
In this approach, natural language is used to explain to the robot, which tasks it should perform.
One such typical task is grasping objects.
Solving this challenge is part of current research such as \cite{karaoguz2019object}, where an object detection approach was used to learn good grasping poses.
Data driven approaches, as stated in \cite{bohg2013data}, are often addressed by providing training data in form of labeled examples, by trial-and-error, or through human demonstration.
This means that communication between human and robot is also important here for a flexible learning progress without prior offline data generation.

\subsection{Joint Attention in Collaborative Settings}
Teaching in collaborative scenarios between human and robot has been investigated by \cite{krause2014learning} and \cite{el2021teaching}, among others.
In~\cite{krause2014learning}, natural language context for one-shot learning of visual objects has been used to enable a robot to recognize a described object.
In this proof of concept, however, the objects had to be unambiguously distinguishable by color or spatial relationships, and the component parts also had to be uniquely describable by linguistic expressions.
In~\cite{el2021teaching}, a teaching system for object categorization was proposed.
This system allowed the user to visualize the intermediate states of categorization, that is, to which category the robot would assign an object.
Through interaction, the categorization could be improved and corrected, but all objects had to be marked with AR markers in order to be recognized at all.
In combination with picking tasks, \cite{valipour2017incremental} and \cite{dehghan2019online} also taught a robot new objects.
In both cases, however, exactly the same objects of each class were used for both training and testing, which positively biases the results.
We use several different objects per class, which is more in line with real-world conditions.

%\url{https://arxiv.org/pdf/2008.00819.pdf}
%\url{https://journals.sagepub.com/doi/pdf/10.5772/5664}

%In agricultural applications for example, the cooperation between human and machine led to a reduction in the amount of pesticides required \cite{berenstein2017human}. This has improved sustainability and reduced the impact on the environment.

In this work, we %build on existing research. We
combine the different research areas of eye tracking and AR for a simple and natural interaction between robot and human to jointly solve a computer vision problem.

	\section{Method} \label{sec:method}
The goal of this work is to teach a robot unknown objects in its environment, in such a way that the robot is later capable of detecting these objects in this same environment.
To this end, we will explain below 1) how we communicate with the robot through the modalities vision, gaze, speech, and gestures 2) how the robot identifies the object of interest, 3) how the human teaches the robot, and 4) how the robot eventually manages to learn.

\subsection{Augmented Reality Interface}
\begin{figure}[t]
    \centering
    \includegraphics[width=\linewidth]{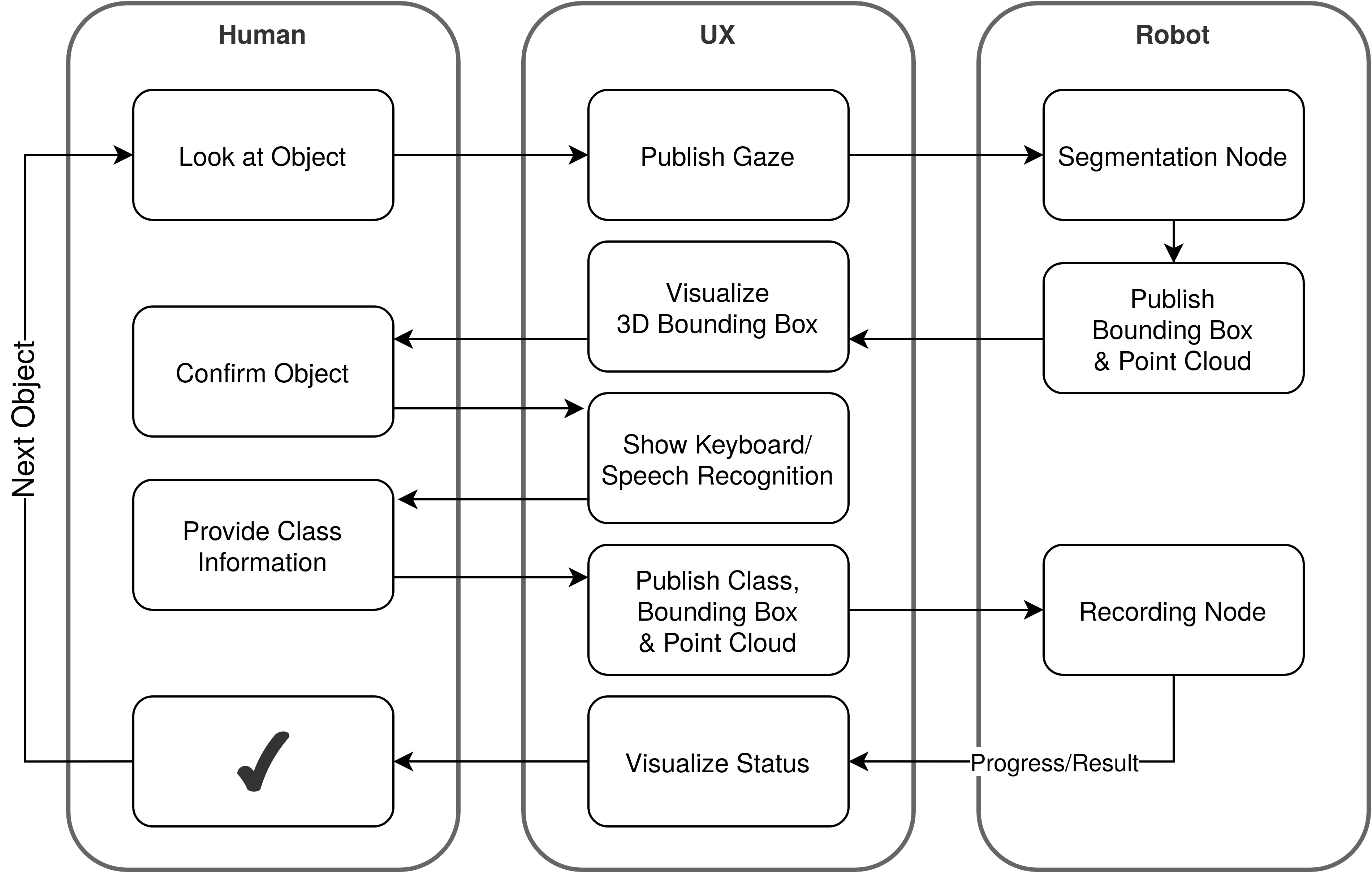}
    \caption{Overview of the entire teaching pipeline. The AR user interface (UX) acts as a bridge between human and robot.}
    \label{fig:interplay}
    \Description{A flow chart showing the intermediate steps of the learning pipeline, each belonging to one of the three main categories, human, user interface or robot. It starts with the human looking at an object (Human category). Then the individual components are processed in the following order: "Publish Gaze" (user interface), "Segmentation Node" (robot), "Publish Bounding Box and Point Cloud" (robot), "Visualize 3D Bounding Box" (user interface), "Confirm Object" (human), "Show Keyboard / Speech Recognition" (user interface), "Provide Class Information" (human), "Publish Class, Bounding Box and Point Cloud" (user interface), "Recording Node" (robot), "Visualize Status" (user interface). Eventually, the human can trigger the pipeline again from the beginning with a new object.}
\end{figure}

In order to enable the human to teach the robot anything, a communication channel is mandatory.
As suggested by \cite{weber2022exploiting}, we meet this need in the form of an augmented reality interface.
The entire communication between human and robot takes place via this interface, that is, the robot can be controlled, the gaze information of the human can be transmitted and the respective class information of the objects can be conveyed.
On the human side, we deploy the HoloLens~2, which is a pair of head-mounted augmented reality glasses manufactured by Microsoft with a built-in eye tracker.
We developed the interface, i.e. the HoloLens application, using the game engine and real-time development platform Unity, version~2019.4.36.
In addition, we used assets from the Microsoft Reality Toolkit, MRTK~2.7.2, which Microsoft supplies specifically for this purpose.
The data interconnection between the Universal Windows Platform (UWP) app on the HoloLens and the robot operating system, ROS~\cite{quigley2009ros}, takes place via ROS\#~\cite{bischoff2019rossharp}.
The open-source software library ROS\# exchanges JSON based commands with ROS through the rosbridge\_suite from within Unity applications.
Both the HoloLens and the robot are continuously connected with each other via WiFi, and data, such as the human's gaze information, can be sent and received in real time.
Finally, gestures and speech serve to operate the AR interface and to interact with the robot.
\figref{fig:interplay} illustrates how the AR interface acts as a bridge between human and robot and shows the interplay of the individual components of our teaching pipeline, which we will describe in more detail below.
% \hl{We make the user interface publicly available under BLIND.}

\subsection{Identifying the Unknown Object of Interest}
In order for the robot to learn a new object, it has to identify it as such in the first place.
This is quite a fundamental problem, as it is, in a sense, a chicken-and-egg problem. 
For the robot, it is difficult to detect the object of interest as it does not know it at this point and it is yet to be taught.
Therefore, since the robot must identify the object before it has learned it, the deployment of neural networks is not possible at this point, and determining where the object begins and where it ends is not trivial.
Instead, we want to incorporate the human's gaze information to help the robot locate the target object. 
This means that the human looks at the object, whereupon the robot can distinguish it from the rest of the environment.
For this purpose, we take the approach of \cite{weber2022exploiting} as a basis, who segmented observed objects using human gaze and the point cloud obtained from the depth sensor of the robot's scene camera.
Thereby, a calibration determines the respective position of the robot and HoloLens, and the HoloLens' motion sensors ensure that the mutual position is tracked during human movements.
In addition, the gaze point, that is, the point at which the human is looking, is continuously tracked by the HoloLens and published as a point in 3D space through our user interface using ROS\#.
Thus, the corresponding ROS topic can be subscribed by the ROS system of the robot, which means that the gaze point is known to the robot at all times and can be used for segmentation.
In the first instance of the segmentation described in \cite{weber2022exploiting} a pass through filter and a voxel grid filter are applied to reduce the size of the point cloud.
Subsequently, the ground is extracted using RANSAC and eventually the object is isolated by means of the gaze point and Euclidean clustering.
We adapt this method with slight adjustments in the last step.
Instead of assigning only the cluster closest to the gaze point to the object, we consider all clusters within a certain distance and with a certain size.
We set the threshold for the maximum distance between cluster and gaze point to 2\,cm and the minimum cluster size to five points.
This way it is possible to even segment very flat objects that do not protrude far from the ground.
Depending on which object the person is currently looking at, the respective object is then segmented in real time.
All the mentioned point cloud processing is accomplished using the open-source Point Cloud Library (PCL) \mbox{\cite{rusu20113d}}.
% \hl{We make the implementation of the object segmentation publicly available under BLIND.}

Owing to the calibration carried out at the beginning, the position of the object of interest is known both from the location of the human wearing the HoloLens as well as from the location of the robot.
The former allows the robot to display its feedback regarding the segmented object as a 3D bounding box on the HoloLens, namely in the human's field of view, using a subscriber, attached to a virtual bounding box, that updates the position of the box depending on the segmentation results.
We can take advantage of the latter during the teaching process described below.

\subsection{Teaching through Joint Attention}
\begin{figure}[t]
    \centering
    \includegraphics[width=\linewidth, trim={0cm 0 0cm 0}, clip]{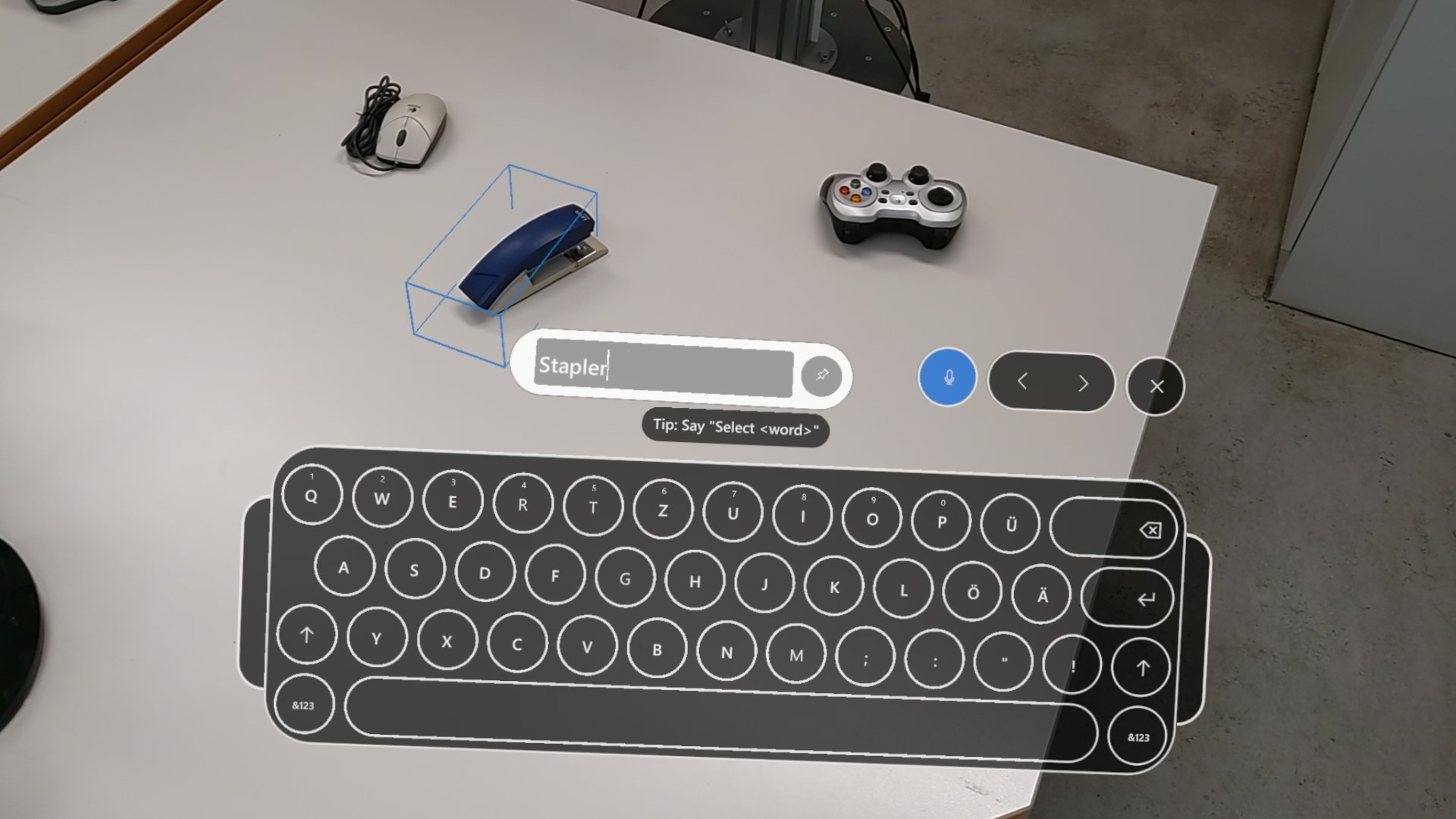}
    \caption{In the human's field of view, a bounding box of the object segmented by the robot is displayed. After the selection it is possible to specify the class name of the respective object using a virtual keyboard or speech.}
    \label{fig:keyboard}
    \Description{Augmented reality interface displayed in the human's filed of view. In the background, various objects lie on a table. A bounding box is shown around a stapler, which is visible to the human via augmented reality. In the foreground, a virtual keyboard can be seen, through which the class name "stapler" has been specified.}
\end{figure}

The attention of the robot and the human is now jointly directed at one and the same object.
The next step is to confirm to the robot that the framed object is the object of interest and to provide a class information.
By performing a pinching gesture with the index finger and thumb (within the field of view of the HoloLens) during the fixation of the object with the human eyes, we can select the object via the AR interface.
Alternatively, it is also possible to just say ``select''.
Thereupon, a virtual keyboard appears in the human's field of vision, on which he can now enter the class name of the object.
Again, it is alternatively possible to simply resort to speech.
\figref{fig:keyboard} shows the implemented keyboard from the human perspective.

\begin{figure}[t]
    \centering
    \includegraphics[width=0.49\columnwidth, trim={23cm 0 14cm 0}, clip]{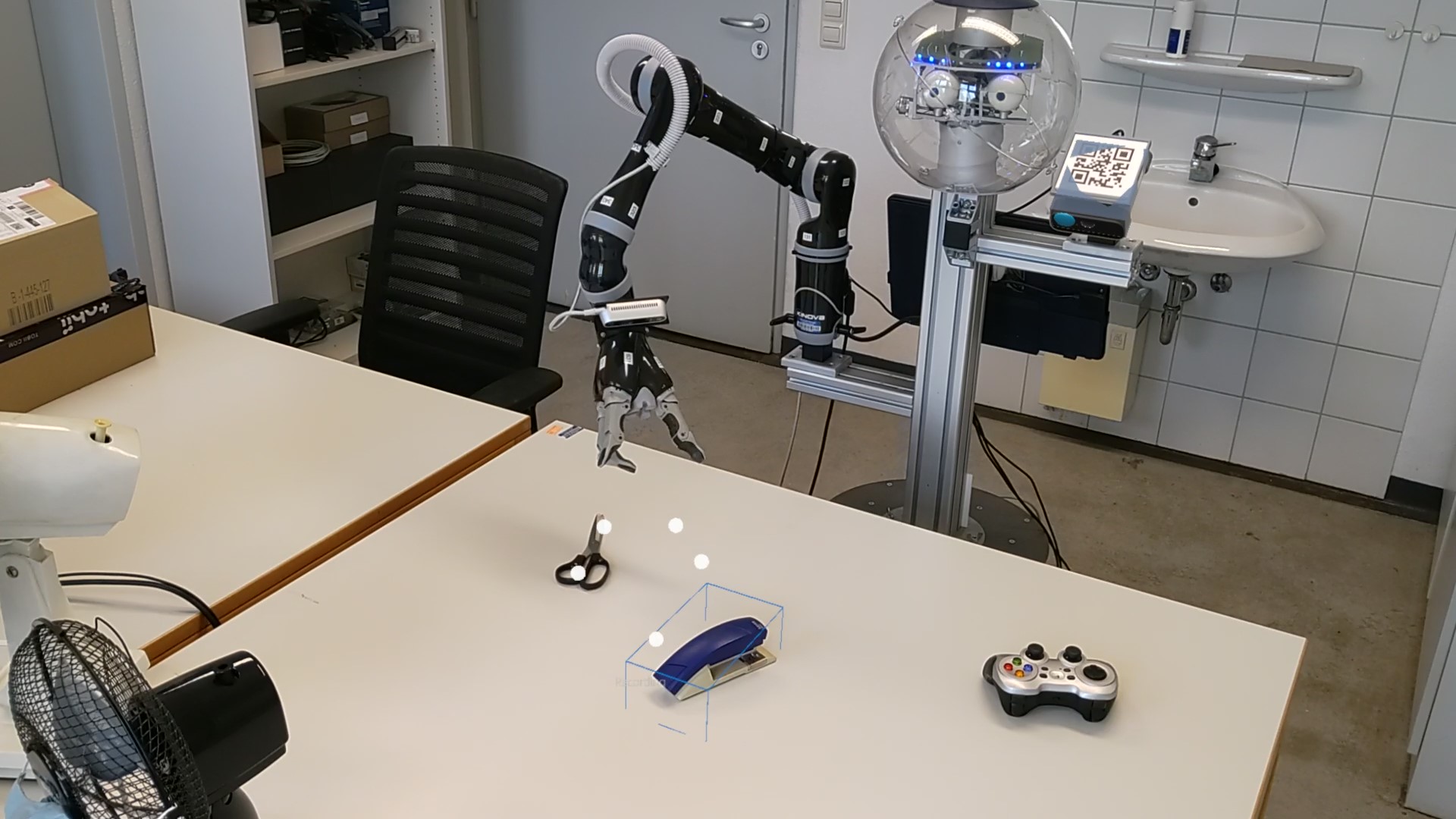}
    \includegraphics[width=0.49\columnwidth, trim={18cm 7.8cm 26cm 6cm}, clip]{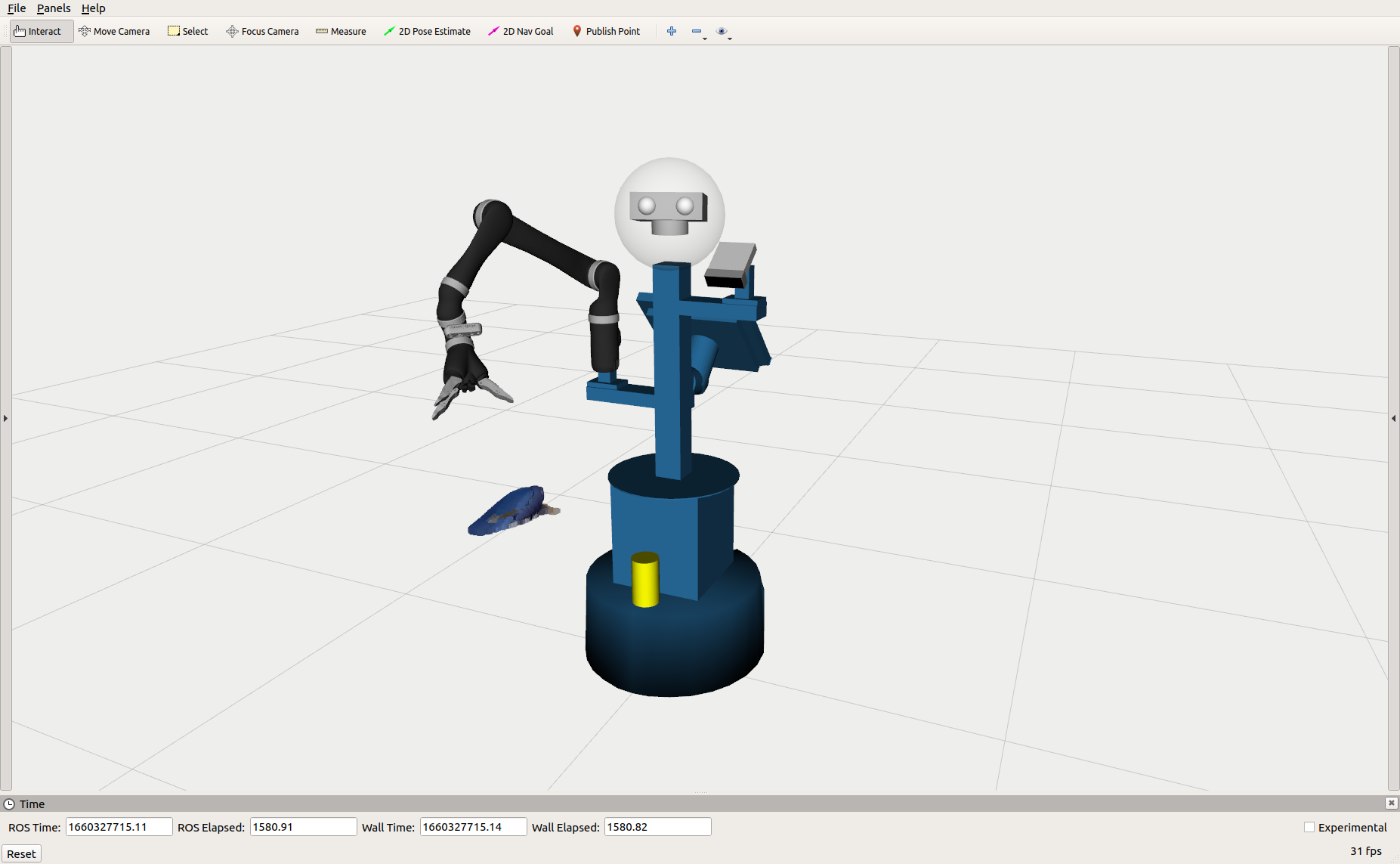}
    \caption{The left side shows the recording process from the human's augmented view and the right side is a visualization in RVIZ with the point cloud of the segmented object. The point cloud is used both to display the bounding box for the human and to label the images captured by the camera on the wrist of the robotic arm.}
    \label{fig:robotarm}
    \Description{Two images of the recording process of an object. On the left side is a robot with a robotic arm, on whose wrist a camera is attached. The arm is outstretched and the camera is pointed obliquely from above at a stapler on the table. The whole scene is shown from the human's field of view and via the augmented reality interface a bounding box can be seen around the stapler, as well as an progress indicator for the ongoing recording process. On the right side the situation is visualized in RVIZ. Instead of the table with all objects only the pointcloud of the stapler is visible.}
\end{figure}

Our next goal is to have the robot autonomously capture images of the object of interest, which it can later use as training data.
In order to get as many images from multiple angles as possible, we attach a second camera to the wrist of the robot's arm.
The robot is now supposed to move this camera in a circle around the object.
This means that once the human has transmitted the class name to the robot by means of a ROS action, it calculates a circular trajectory of reachable points.
Due to physical limitations, such as the length of the arm, this is usually a partial segment of the circle.
During the movement, the camera is aligned in such a way that it points at a 45 degree angle to the center of the previously determined 3D bounding box.
As the distance between the center and the camera, we use twice the length of the diagonal of the bounding box with a minimum safety distance of twice the distance between the camera and the robot arm end effector.
The recording process is illustrated in \figref{fig:robotarm}.

By virtue of the calibration described in the previous section, the robot is not only aware of the position of the object of interest, but also capable of computing the transformation to all of its coordinate systems, namely the robot frames.
This includes the camera on the robot's arm.
The essential aspect in this step is to continuously transform the point cloud of the segmented object into the coordinate system of the camera.
By projecting the point cloud onto the 2D image plane of the camera, we can derive the region of interest from the boundary points.
Thus, for each captured image, we can additionally store a 2D bounding box calculated in this way.
Overall, the stored synchronized data are RGB and depth images as well as the regions of interest with the 2D bounding boxes.
In addition, we also store the positions of the camera to the captured object.
%\begin{algorithm}[hbt]
%    \caption{An algorithm with caption}\label{alg:cap}
%    \begin{algorithmic}[1]
%        \Require class, object
%        \State $W \gets \operatorname{calcCircularTrajectory(center, radius)}$
%        %\State $W \gets \operatorname{checkReachable(W)}$
%        %\State $T \gets \operatorname{calcPath(W)}$
%        \State execute(W)
%        \While{moving}
%            \State pc3D' $\gets \operatorname{transform(pc3D)}$ 
%            \State pc2D $\gets \operatorname{project3DToPixel(pc3D')}$ 
%            \State roi $\gets \operatorname{min/max(pc2D)}$
%            \State Store: RGB, depth, TF, roi
%        \EndWhile
%    \end{algorithmic}
%\end{algorithm}
Eventually, the robot is able to automatically produce hundreds of labeled training images in a remarkably brief period of time.
More precisely, teaching one object takes about one minute and yields about 300 images.
% More precisely, around 300 images per taught item.
The progress and completion of the recording process is in turn transmitted to the HoloLens via the ROS action, visualizing to the human when the robot is ready for a new object.
% \hl{We make the implementation of the recording node publicly available under BLIND.}

\subsection{Transfer Learning}
Following the teaching part, the learning process of the robot now ensues.
To enable the robot to independently detect the previously seen objects in the future, it must use the information at its disposal in the form of the training data it has created itself.
In other words, the robot, or rather its neural network based object detectors, will be trained on the RGB images obtained.
This will be accomplished by means of transfer learning.
Hence, we assume some prior awareness of objectness, since our method should be seen as an extension rather than a replacement for the training with common large datasets, such as ImageNet~\cite{deng2009imagenet}, PASCAL~VOC~\cite{everingham2010pascal} or MS~COCO~\cite{lin2014microsoft}.
Such an approach is realistic, since most of the existing objects are not part of these datasets.
We aim to extend this state of knowledge with our method.
Consequently, we resort to state-of-the-art object detectors, such as Faster \mbox{R-CNN}~\cite{ren2015faster} and FCOS~\cite{tian2019fcos}.
Starting from one of these pretrained models respectively, we delete the last classification layers, and then reinitialize them with the appropriate number of output neurons for our use case.
Finally, we retrain said layers with our data, freezing all other neurons from the preceding feature layers.
By freezing the feature layers responsible for the general comprehension of objects and fine-tuning only the last few layers, we prevent overfitting~\cite{yosinski2014transferable, zhuang2020comprehensive}.
In fact, since we only retrain the classification heads of the models, even training on the robot itself is possible without relying on a high-performance GPU.
Naturally, the training may take longer.
In \secref{sec:evaluation}, we will evaluate the performance of the aforementioned models, among others.
% \hl{We make the transfer learning implementation publicly available under BLIND.}

	\section{Dataset: Objects in Multiperspective Detail} \label{sec:dataset}
The method introduced in the previous section allowed us to create a new type of dataset.
While we publish the validation and test set mainly for the sake of reproducibility of our results, we think that our training set might be especially interesting for further purposes.
In the following, we would like to explain the training data in more detail and provide some statistics.
We will elaborate on the validation and test set within the scope of our evaluation in \secref{sec:evaluation}.

Our training set is particularly characterized by the fact that it supplies many details about individual objects.
While most object detection datasets often consist of many images with different objects, our data depicts the objects from many different angles.
This is especially attractive for objects that appear different from the front and back, such as a gamepad.

\begin{figure}[t]
    \centering
    \includegraphics[width=0.49\columnwidth, trim={0cm 0 0cm 0}, clip]{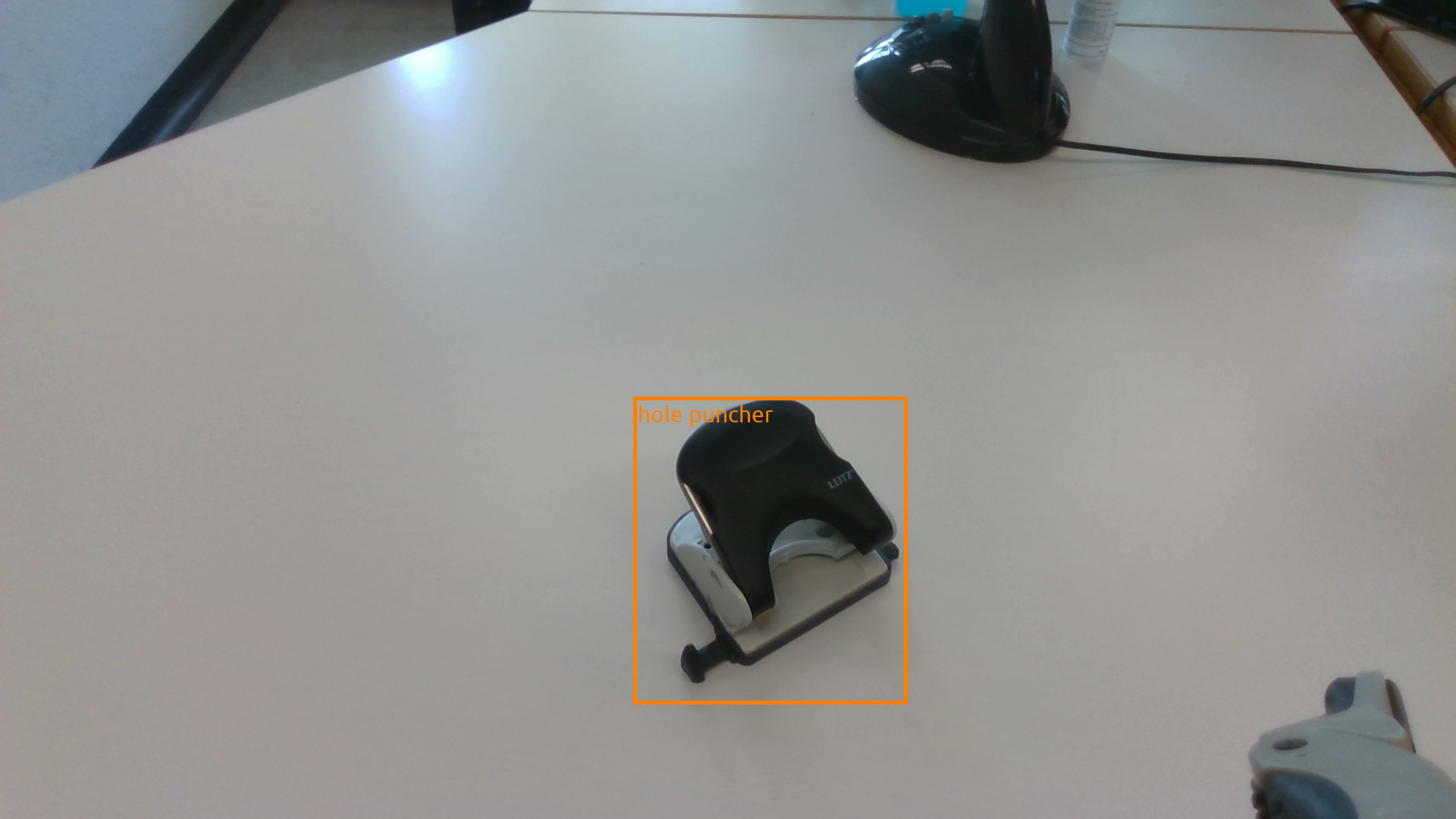}
    \includegraphics[width=0.49\columnwidth, trim={0cm 0 0cm 0}, clip]{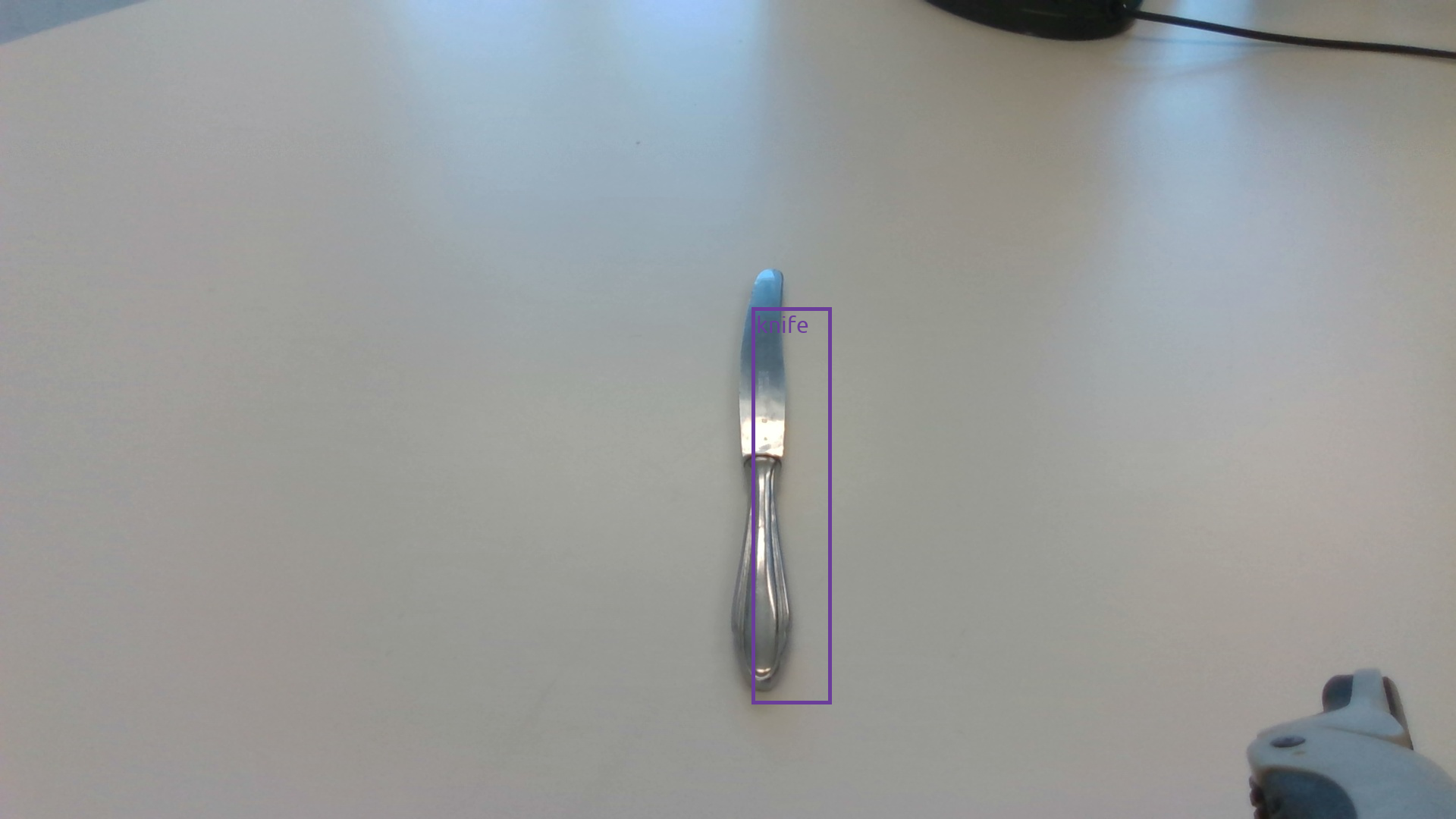}
    \par \vspace{2pt}
    \includegraphics[width=0.49\columnwidth, trim={0cm 0 0cm 0}, clip]{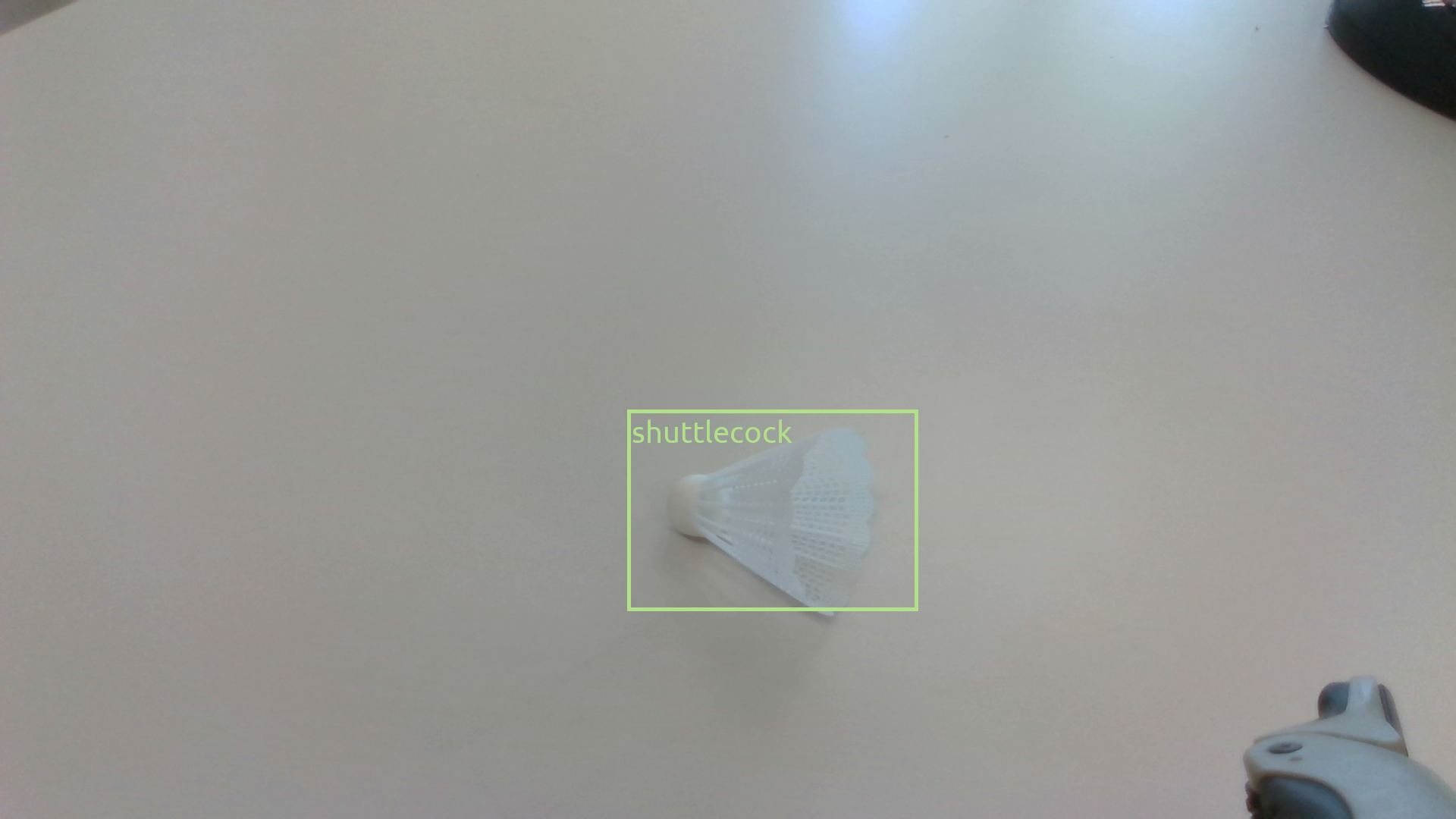}
    \includegraphics[width=0.49\columnwidth, trim={0cm 0 0cm 0}, clip]{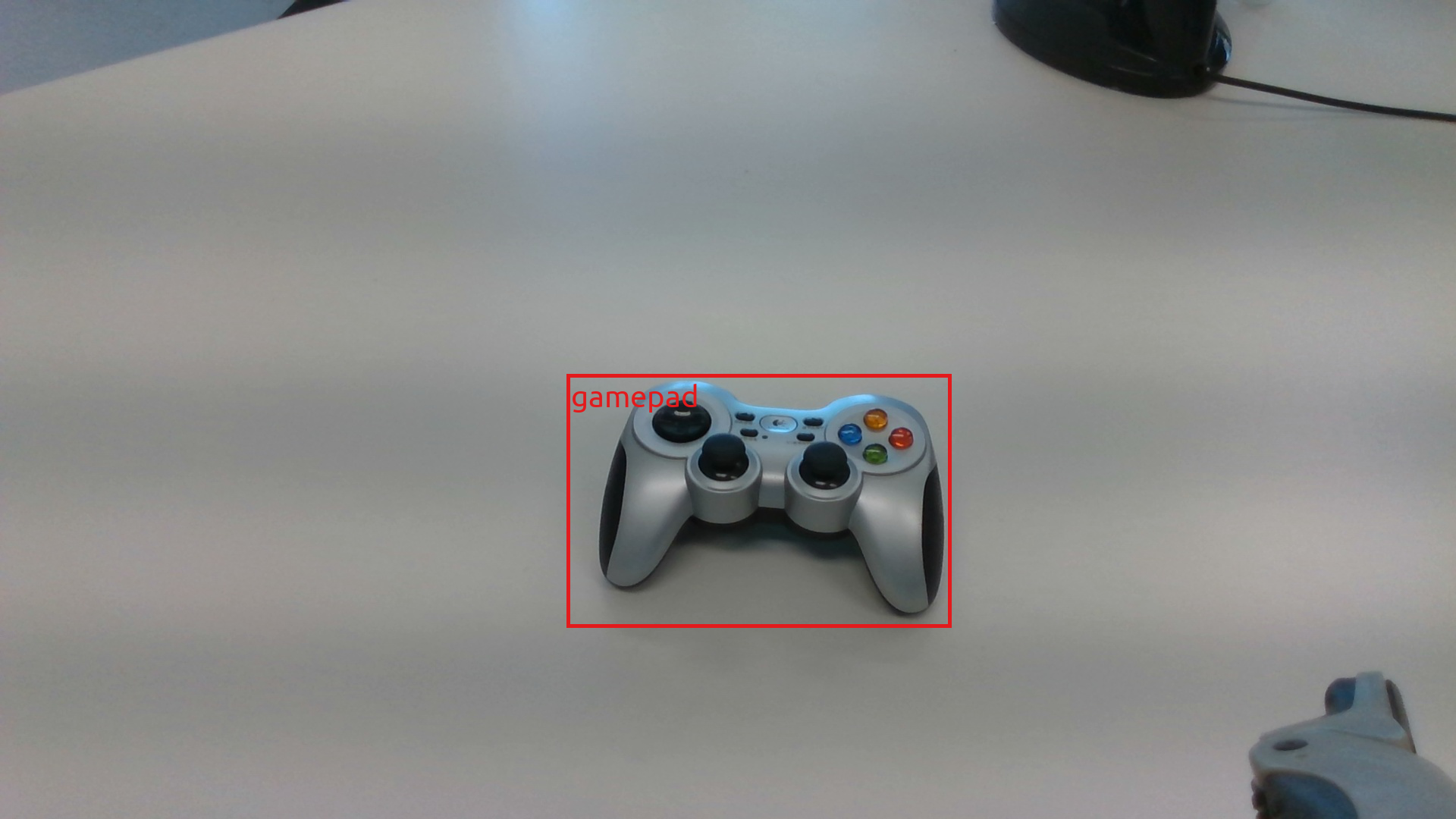}
    \caption{Sample images of a hole puncher, a knife, a shuttlecock and a gamepad from our dataset. The quality of the bounding boxes may vary depending on the point of view and may sometimes be slightly too large, too small or offset. In all images, however, the majority of the box always covers the respective object. The objects contrast differently with the background in terms of flatness and color.}
    \label{fig:dataset}
    \Description{Four images of four different objects on a table. Around each object is drawn the bounding boxes determined by the robot. The bounding box around the hole puncher and the gamepad are very fitting, while the box around the slim fork has a minimally offset and the one around the shuttlecock is slightly too large.}
\end{figure}

\begin{figure}[!t]
    \centering
    \includegraphics[width=.75\linewidth]{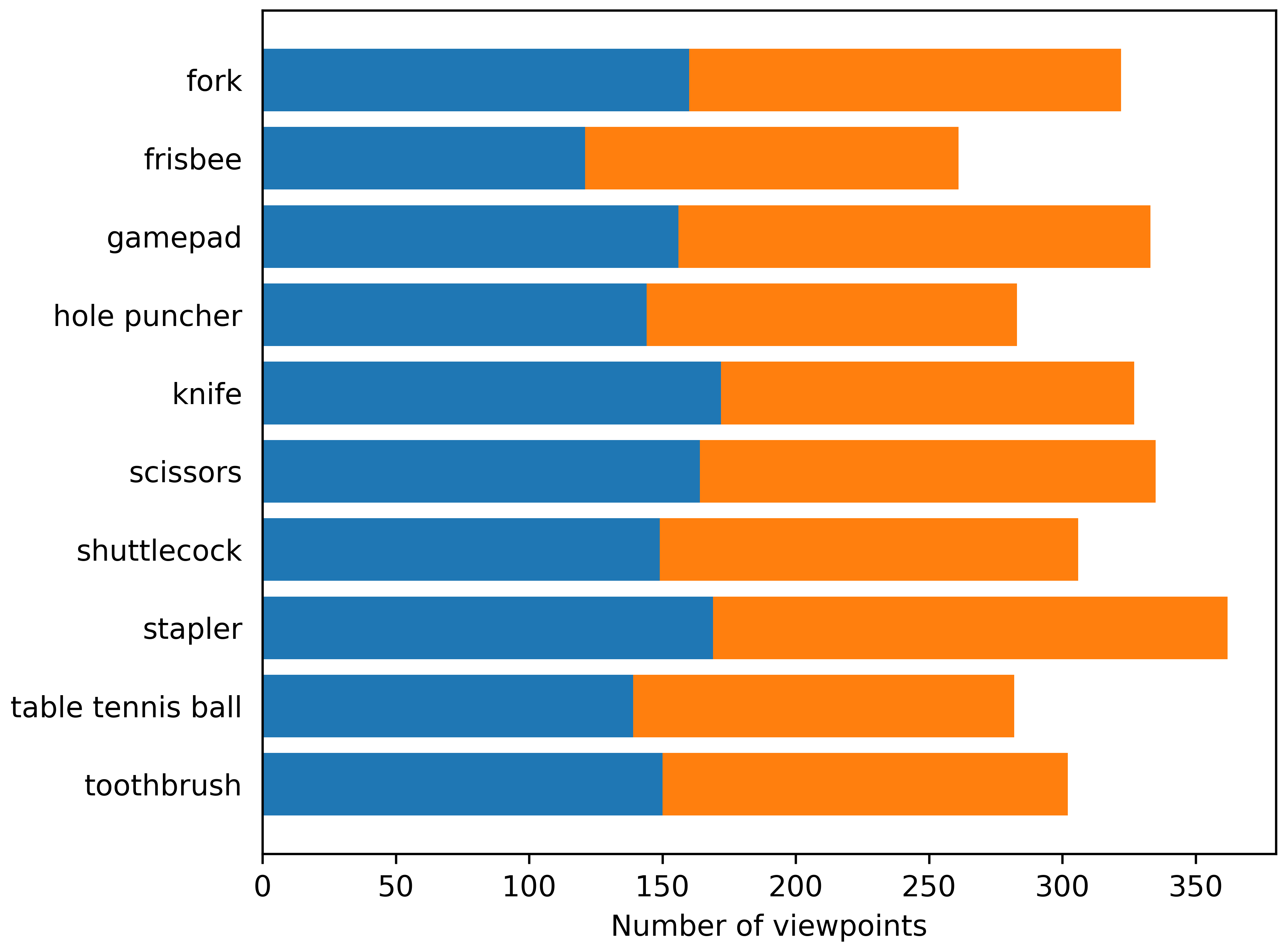}
    \caption{Distribution of the viewpoints across the categories. The colors indicate the two different items within the classes.}
    \label{fig:num_classes}
    \Description{A bar chart showing the number of viewpoints for each class in the dataset. The ten classes are listed on the y-axis and the number of viewpoints on the x-axis. All classes range from 280 to 380 viewpoints. The class frisbee has the fewest viewpoints and the class stapler has the most. All ten bars consist of two bars in different colors, which represent the two entities within a class. For all classes, the color change is approximately in the middle, which means that there are approximately the same number of viewpoints of both entities in the dataset.}
\end{figure}

The set consists of 3113 perspectives in total and contains the classes fork, frisbee, gamepad, hole puncher, knife, scissors, shuttlecock, stapler, table tennis ball and toothbrush.
For each class, there are two different entities in the set, differing in color, shape, or both.
\figref{fig:dataset} shows some sample images and \figref{fig:num_classes} illustrates the distribution of the viewpoints.
%\begin{table}[htb]
%    \caption{Characteristics of the bounding boxes in the data set. The numbers represent the respective bounding box size in pixels. Note that the values in the columns can belong to different bounding boxes. The variables $ \mu $ and $ \sigma $ denote the mean and the standard deviation.}
%    \label{tbl:datasetStats}
%    \centering
%    \begin{tabular}{c|rrr}
%        & Width & Height & Size \\
%        \midrule
%        min     & 68    & 43    & 25\,929 \\
%        max     & 689   & 537   & 239\,117 \\
%        $\mu$   & 426   & 318   & 129\,684 \\
%        $\sigma$& 144   & 106   & 50\,058 \\
%    \end{tabular}
%\end{table}
The objects are each placed individually on a table and for every camera perspective multiple pieces of information are included.
For each RGB image, alongside the region of interest, there is a corresponding depth image that is aligned to the color image.
All RGB images and depth images have a resolution of $ 1920 \times 1080 $.
In addition, all camera poses are available.
They are specified independently of the robot as a transformation, composed of translation vector and rotation quaternion, from the coordinate system of the camera to the one of the object.
The last component is the meta-information about the camera with the intrinsic parameters of the camera calibration.

Altogether, we believe that the high information density in our data is also interesting for other research areas where camera positions are crucial, such as Neural Radiance Fields~\cite{mildenhall2020nerf, yu2021pixelnerf, lin2021barf, deng2022depth}.
There, either synthetic data must be used or, given real data, the camera positions (and depths) can only be roughly approximated via structure from motion.
For this reason, we make our dataset, Objects in Multiperspective Detail (OMD), publicly available to the research community.

	\section{Evaluation} \label{sec:evaluation}
For all our experiments, alongside the aforementioned HoloLens 2 worn by the human, we employed a Scitos G5 from MetraLabs~\cite{MetraLabs} as robot.
The body camera through which the robot observes the scene and performs the segmentation is an Azure Kinect DK from Microsoft.
The robot arm that was additionally installed on the Scitos is a Kinova Jaco2~\cite{KinovaGen2} with 6 DoF.
The camera attached to the wrist of the arm in order to take pictures of the objects is an Intel RealSense D435.

\begin{figure}[hb]
    \centering
    \includegraphics[width=0.49\columnwidth, trim={0cm 0 0cm 0}, clip]{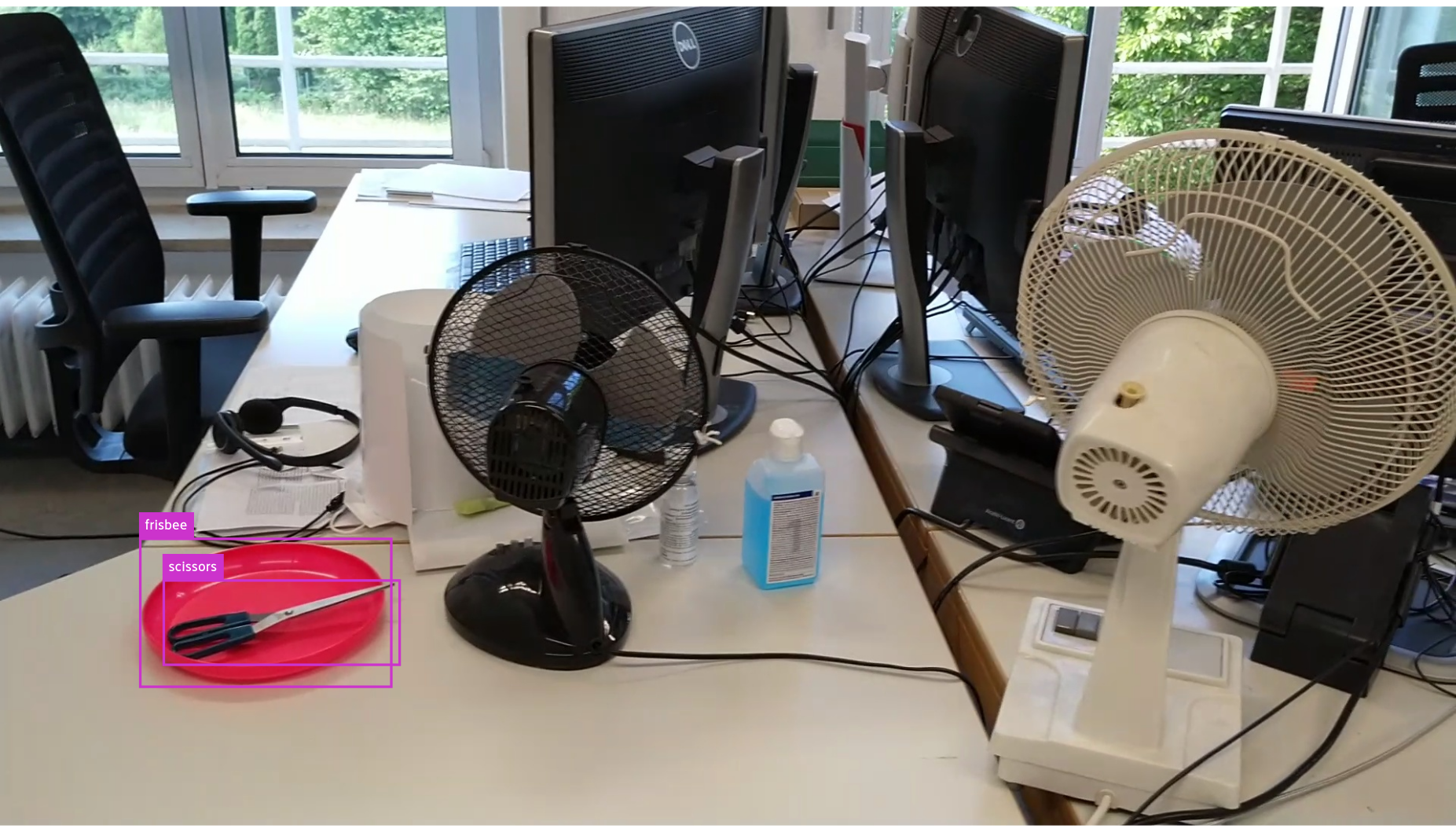}
    \hfill
    \includegraphics[width=0.49\columnwidth, trim={0cm 0 0 0}, clip]{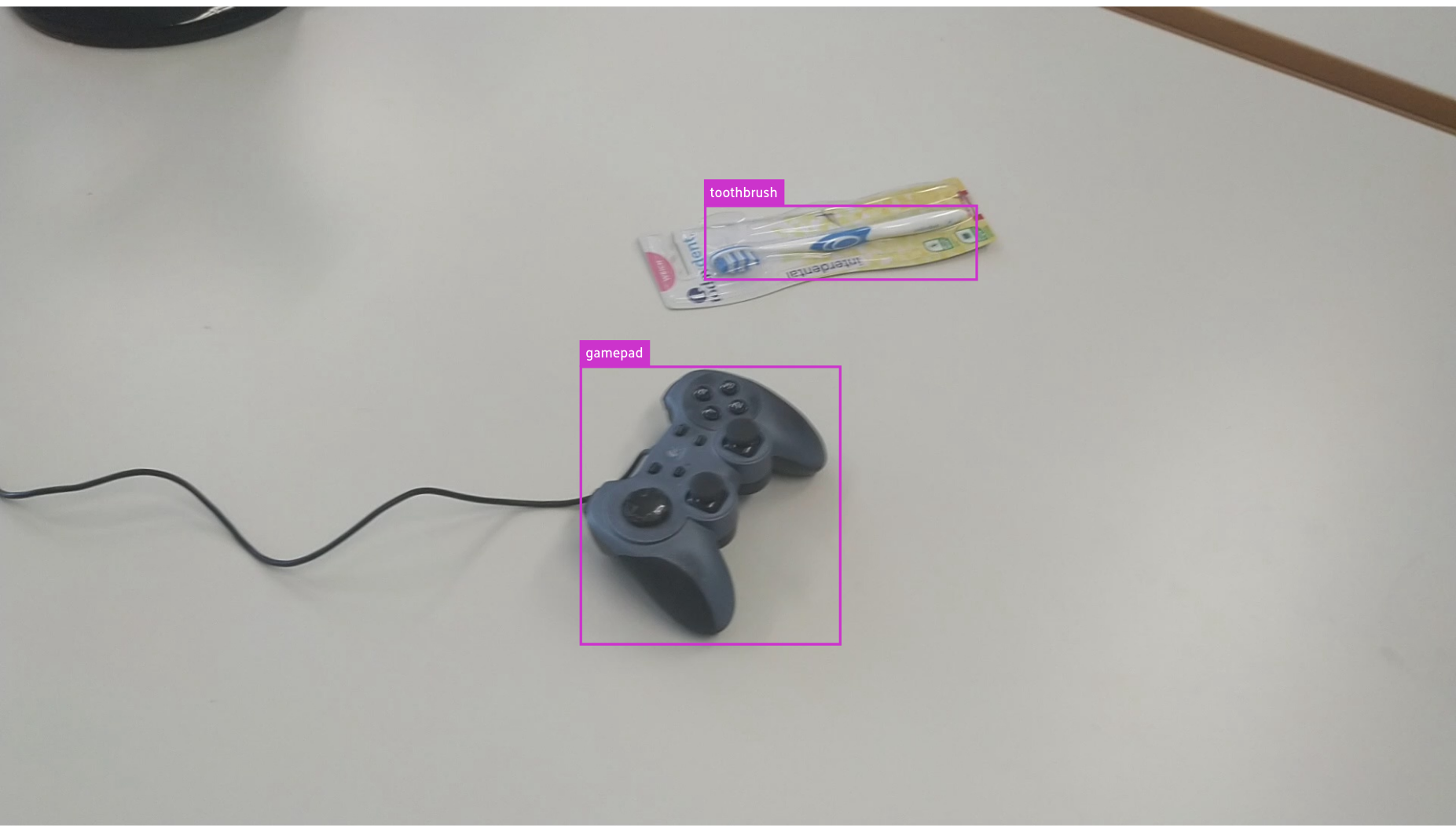}
    \caption{Sample images from the test set. The set of objects is disjoint with the ones from the training set (see gamepad). The set is also diverse in terms of the clutter of the background and the distances to the objects.}
    \label{fig:dataset_test}
    \Description{Two images of objects in an office environment. On the left is a pair of scissors in an upside-down frisbee surrounded by office items, monitors, fans and other things. On the right side a gamepad and a toothbrush are lying on a white table. The toothbrush is still in its packaging.}
\end{figure}

For training we use our own training set, which we have explained in detail in \secref{sec:dataset}.
Since the objective of this work is to teach the robot its environment, we also had to record a validation and test set located in the environment where the learning process took place.
As mentioned earlier, alongside the training set, we will also publish the validation and test set to ensure the reproducibility of our results.
The validation and test set consist of 1051 and 1410 regular images, respectively, which were manually labeled by hand using DarkLabel~\cite{DarkLabel}.
The classes represented therein are the same ten as in the training set.
For each class, four distinct objects of the respective category were available.
The validation set contains the same two objects of each class as the training set, whereas the test set contains the other two.
That is, the objects differ in shape, color, or both from those used in training.
The objects were photographed randomly grouped (within the set) in the robot's office environment.
We made sure to create challenging scenarios as well, such as items being stacked or the toothbrush still being in its packaging, as shown in \figref{fig:dataset_test}.

In the following, we evaluate our learning pipeline using several state-of-the-art object detectors.
Consequently, the object detectors Faster \mbox{R-CNN}~\cite{ren2015faster}, RetinaNet~\cite{lin2017focal}, FCOS~\cite{tian2019fcos}, and SSD300~\cite{liu2016ssd} serve as a foundation.
We complement these with various backbones, such as ResNet-50-FPN~\cite{he2016deep}, VGG16~\cite{simonyan2014very} and MobileNetV3 Large~\cite{sandler2018mobilenetv2, howard2019searching}.
All backbones were trained on ImageNet and can be left as is, since we deliberately picked object classes for our evaluation that had no intersection at all with this dataset.
The reason behind our choice of the ten test objects was as follows.
On the one hand, they must not appear in ImageNet due to the backbones, but on the other hand, at least a part of them ought to be in MS COCO so that we have a comparison later on.
Furthermore, within each class there had to be several different looking objects of that class.
All of this together limited the selection accordingly, especially since we tried to avoid perishable classes like food.
Hence, in terms of the actual object detectors and to ensure that our objects are indeed unknown to the models, we had to train them on a subset of MS COCO.
More precisely, we extracted the classes fork, frisbee, knife, scissors, sports ball, and toothbrush from the dataset using the tool Fiftyone~\cite{moore2020fiftyone} and then trained the above mentioned detectors on the remaining part.
In doing so, we followed the respective training recipe of the original implementation and, for consistency, adhered thereto in all of our subsequent experiments in our own training pipeline.
%Following the pretraining, we executed our own training pipeline.
The only exception in our transfer learning approach was the type of data augmentation applied.
In this case, we used random photometric distortion, random zoom out, random cropping, and random horizontal flipping for all models (not just SSD300) to prevent overfitting.
Subsequently, we trained using the method described in \secref{sec:method}.

In all our experiments, we evaluate according to the MS~COCO metric~\cite{lin2014microsoft}, namely the average precision for varying intersection-over-union thresholds (IoU).
In this context, we use the abbreviations $ \AP{} = \text{AP}^{\text{IoU}=0.5:0.05:0.95} $, $ \AP{50} = \text{AP}^{\text{IoU}=0.5} $, and $ \AP{75} = \text{AP}^{\text{IoU}=0.75} $ within a class and \mAP{} as the average over all categories.
Analogously, this applies to the average recall, where we consider the maximum recall given 1, 10, and 100 detections per image, respectively, and use the abbreviations $ \AR{1} = \text{AR}^\text{max=1} $, $ \AR{10} = \text{AR}^\text{max=10} $ and $ \AR{100} = \text{AR}^\text{max=100} $.
Again, \mAR{} denotes averaged over all categories.
Unless otherwise stated, average precision and average recall refer to \AP{} and \AR{}, respectively.

A comparison of all tested models is provided in \tblref{tbl:evalModels}.
\begin{table*}[tb]
    \centering
    \caption{Comparison of all machine learning models trained in a transfer learning fashion. The best values are highlighted in bold.}
    \label{tbl:evalModels}
    \begin{threeparttable}
    \begin{tabular}{cccccccc}
    \toprule
        Model   & Backbone  & \mAP{} & \mAP{50} & \mAP{75} & \mAR{1} & \mAR{10} & \mAR{100}  \\
        \midrule
        Faster R-CNN~\cite{ren2015faster}   & ResNet-50~\cite{he2016deep}
                                            & \textbf{33.6} & \textbf{66.9} & 31.4 & 43.7 & 50.1 & 50.4 \\
        Faster R-CNN~\cite{ren2015faster}   & MobileNetV3~\cite{sandler2018mobilenetv2,howard2019searching} 
                                            & 15.5 & 38.1 & \pz6.1 & 23.7 & 27.4 & 27.7 \\
        Faster R-CNN~\cite{ren2015faster}   & MobileNetV3~\cite{sandler2018mobilenetv2,howard2019searching}\tnote{$\dagger$}
                                            & 13.0 & 38.4 & \pz3.1 & 22.2 & 25.5 & 25.5 \\
        FCOS~\cite{tian2019fcos}            & ResNet-50~\cite{he2016deep}
                                            & 30.6 & 47.6 & \textbf{35.9} & 44.7 & 53.8 & 55.0 \\
        RetinaNet~\cite{lin2017focal}       & ResNet-50~\cite{he2016deep}
                                            & 31.2 & 52.4 & 34.3 & \textbf{46.2} & \textbf{57.6} & \textbf{59.1} \\
        SSD300~\cite{liu2016ssd}            & VGG16~\cite{simonyan2014very}
                                            & \pz8.0 & 19.1 & \pz5.0 & 21.0 & 31.6 & 34.0 \\
    \bottomrule
    % \multicolumn{2}{l}{$^\dagger$ \footnotesize Tuned for mobile use cases}
    \end{tabular}
    \begin{tablenotes}
			\item[$\dagger$] \footnotesize Tuned for mobile use cases
	\end{tablenotes}
    \end{threeparttable}
\end{table*}
Faster \mbox{R-CNN} with the ResNet-50 backbone generally performed best in terms of average precision.
With a MobileNetV3 backbone, the performance was significantly worse in terms of both precision and recall. 
FCOS and RetinaNet are slightly behind Faster \mbox{R-CNN} in terms of the mean average accuracy.
The latter has the best recall values, while FCOS has the best mean average precision at an intersection-over-union of $ 0.75 $.
SSD300 clearly lags behind all other models in terms of precision.

\begin{figure}[!t]
    \centering
    \includegraphics[width=.7\linewidth, trim={0cm 0 0 0}, clip]{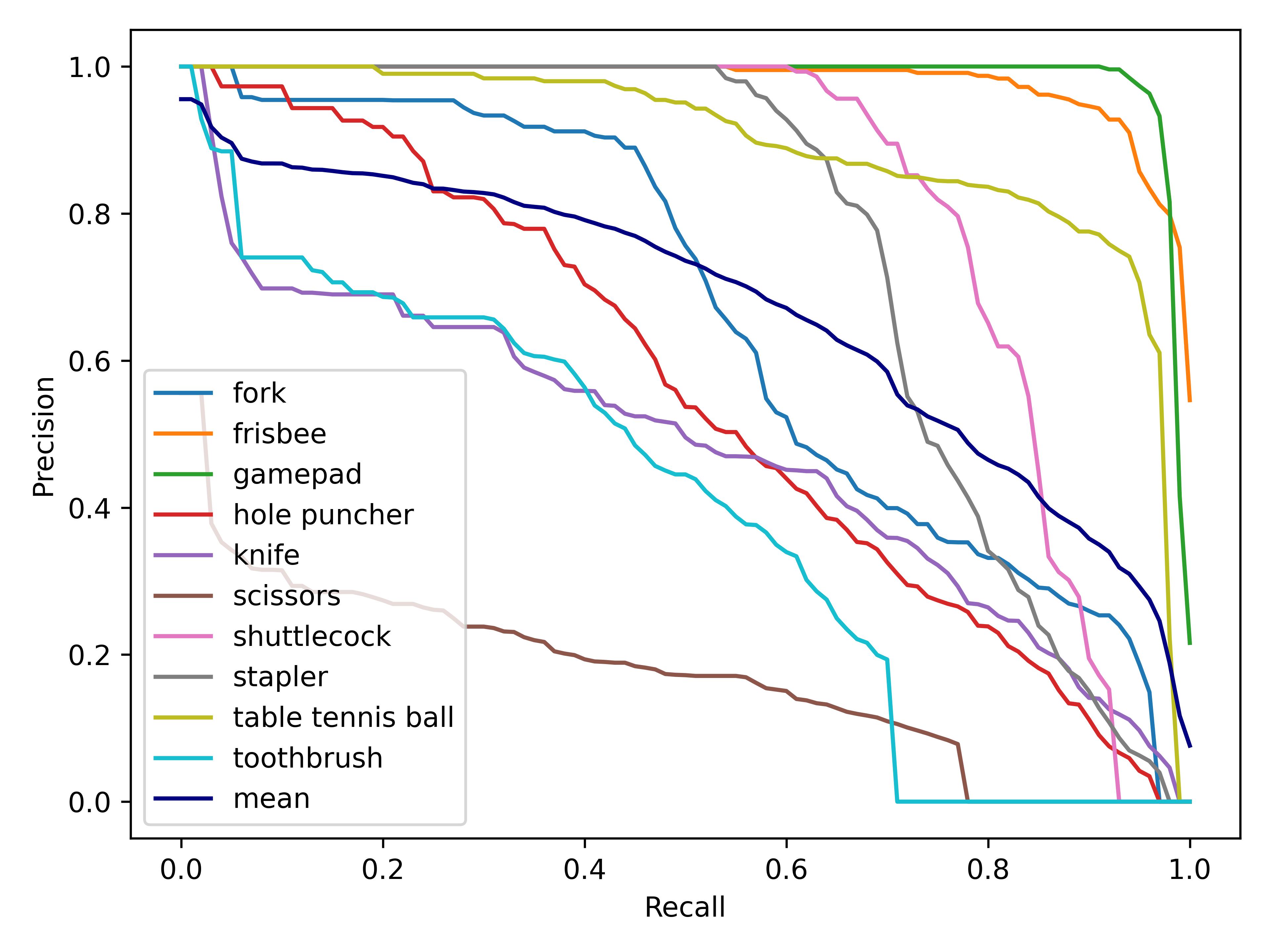}
    \caption{Precision-recall curve of Faster \mbox{R-CNN} at an IoU of~0.5. Above this value, objects can be considered as detected~\cite{everingham2010pascal, zitnick2014edge}.}
    \label{fig:PRCurve}
    \Description{A line graph of the precision recall curves of the ten classes. The recall is shown on the x-axis and the precision on the y-axis. The curves start at a recall of 0.0 with precision values in the range of 1.0 and decrease with increasing recall. The curve of the scissors is the only one that starts in the range of 0.4 and the precision decreases evenly until a recall of 0.8 to zero. The toothbrush reaches zero a bit earlier, but is overall far above the curve of the scissors, since it starts at a precision of 1.0.  The gamepad and the frisbee are still almost at a precision of 1.0 up to a recall of 0.9. All other curves lie between the toothbrush and the gamepad.}
\end{figure}

As we proceed, we will continue with Faster R-CNN for further analysis, since MS~COCO~\cite{lin2014microsoft} considers \mAP{} as the single most important metric.
In general, the model seems to detect the objects quite well, but some classes cause more difficulties than others.
This also becomes apparent by looking at the curve in \figref{fig:PRCurve}.
Even at low recall values, the precision of the scissors is below $ 0.5 $.
The class toothbrush also decreases early.
On the other hand, the precision for the classes gamepad and frisbee is consistently excellent, even for a high recall.

In \tblref{tbl:evalClasses}, we compare different training variants.
\begin{table}[!b]
    \centering
    \caption{Comparison of the average precision (\AP{}) for different training types of Faster \mbox{R-CNN} on our test set. Namely, apart from the backbone, trained from scratch (S) or trained in the sense of transfer learning with frozen non-classification layers (TL-F) or completely unfrozen (TL-U), respectively. All three on the data collected by the robot. The best values are highlighted in bold. The last column (COCO) serves as an orientation and reports the results of Faster \mbox{R-CNN} trained on the entire MS COCO training set.}
    \label{tbl:evalClasses}
    % \begin{threeparttable}
    \begin{tabular}{rccc|c}
    \toprule
        Class               & S     & TL-U  & TL-F  & COCO \\
        \midrule
        fork                & \textbf{21.7}  & \textbf{21.7}  & 18.6  & 63.8 \\
        frisbee             & 19.5  & 47.6  & \textbf{58.8}  & 65.9 \\
        gamepad             & 38.4  & 24.8  & \textbf{62.6}  & -    \\
        hole puncher        & \textbf{42.5}  & 26.3  & 23.0  & -    \\
        knife               & 20.4  & 19.1  & \textbf{27.6}  & 50.0 \\
        scissors            &\pz6.5 &\pz\textbf{7.3} &\pz5.1 & 70.9 \\
        shuttlecock         & 24.9  & 27.8  & \textbf{51.5}  & -    \\
        stapler             & 29.3  & 24.8  & \textbf{38.9}  & -    \\
        table tennis ball   &\pz3.4 & 27.6  & \textbf{44.0}  & 17.3 \\
        toothbrush          & \textbf{21.6}  & 8.8   & 6.2   & 37.4 \\
        \midrule
        \mAP{50}            & 62.8  & \textbf{68.8}  & 66.9  & 84.4 \\
        \mAP{75}            &\pz9.8 & 7.5   & \textbf{31.4}  & 55.4 \\
        \mAP{\phantom{00}}  & 22.8  & 23.6  & \textbf{33.6}  & 50.9 \\
    \bottomrule
    \end{tabular}
    % \end{threeparttable}
\end{table}
If we ignore the baseline variant (COCO) for a moment, we find that the transfer learning method, in which only the last layers had been trained (TL-F), has the best \AP{} for the majority of classes.
In particular, \mAP{75} is significantly higher.
It is worth mentioning that the baseline values (COCO) are naturally superior.
The difference between the full MS COCO training set and the part we used for pretraining remains still $ 25\,713 $ objects in $ 14\,296 $ images, which is five times as many images as we used.
In addition, our images are distributed among all ten classes, while MS COCO does not contain four of them and the baseline thus does not recognize them at all.
This demonstrates the strength of our pipeline, which is designed to enable the learning of additional, as yet unknown classes, that is, to extend existing knowledge.
The comparison with the baseline, which is the ideal case, namely 1) a suitable data set exists 2) it is accessible and 3) the object is part of it, serves primarily to better classify our results into the overall picture.
It is not intended to outperform a model trained on such a large data set, but rather to determine an upper bound and test how close we can get with our method.
Taking this into account, it is remarkable how well our pipeline has learned especially the classes gamepad or shuttlecock, whose \AP{} is even higher than the \mAP{} of the baseline.
Furthermore, since the table tennis ball occurs in MS COCO only as a subset of the class sports ball, we can see how our system becomes more attentive to table tennis balls.
In contrast, the class frisbee does not quite reach the baseline as the corresponding MS COCO class contains exclusively frisbees.
In the case of the category hole puncher, the results are satisfying even without pre-existing basic knowledge (S).
% categories fork, hole puncher and toothbrush

Considering the amount of time needed for the respective training, major differences become apparent.
Training of the entire MS COCO training set lasted the longest, at more than two days on two NVIDIA RTX A4000 deployed in parallel.
The other three variants were trained on our dataset on only one of the GPUs and took 4 hours for the entire model (S, TL-U) and 2.5 hours for the freezed variant (TL-F), respectively.
As mentioned above, although for consistency reasons we trained 26 epochs as in the original recipe, the weights with the best validation accuracy that we eventually used for testing were often reached earlier.
For Faster \mbox{R-CNN} trained via TL-F, this was even the case after three epochs (starting at $ \mAP{}=0.0 $ before training), which corresponds to a training time of about 40 minutes on our Scitos equipped with an NVIDIA GeForce GTX 1050 Ti.
This makes the entire pipeline also suitable for stand-alone learning directly on the robot.

Finally, we analyze the influence of the amount of images used for training.
\tblref{tbl:evalSubset} lists the results of Faster R-CNN trained using TL-F for varying dataset sizes.
\begin{table}[tb]
    \centering
    \caption{Results of Faster \mbox{R-CNN} trained via TL-F on different sized subsets of our dataset. The best values are highlighted in bold.}
    \label{tbl:evalSubset}
    % \begin{threeparttable}
    \begin{tabular}{ccccc}
    \toprule
                            & 25\%  & 50\%  & 75\%  & 100\%\\
        \midrule
        \mAP{\phantom{000}} & 31.5  & 32.9  & 31.7  & \textbf{33.6} \\
        \mAP{50\phantom{0}} & 64.7  & 66.7  & 66.7  & \textbf{66.9} \\
        \mAP{75\phantom{0}} & 28.6  & 28.4  & 26.0  & \textbf{31.4} \\
        \mAR{1\phantom{00}} & 41.1  & 42.9  & 41.4  & \textbf{43.7} \\
        \mAR{10\phantom{0}} & 46.9  & 49.9  & 46.9  & \textbf{50.1} \\
        \mAR{100}           & 47.2  & 50.2  & 47.1  & \textbf{50.4} \\
    \bottomrule
    \end{tabular}
    % \end{threeparttable}
\end{table}
The images were removed from the sequence of perspectives with equidistant spacing.
Apart from the case where 75\% of the data is used, the tendency emerges that as the number of images increases, so does the average precision and average recall.
The best result is obtained using all the data.

%\todo{Other eval ideas see \url{https://www.mdpi.com/1424-8220/22/4/1352/pdf}, \eg confusion matrix}

	\section{Limitations \& Discussion}
Similar to all supervised machine learning methods, we depend on the quality of the training data.
In our case, this can vary depending on the preceding segmentation of the point cloud.
This in turn is naturally dependent on the quality of the data obtained by the depth sensor of the robot's scene camera.
Especially with very dark or glossy surfaces, we noticed that the depth sensor had problems determining the depth accurately.
As a result, the accuracy of the bounding boxes suffers, which eventually has an impact on performance.
However, this problem can be compensated with an even larger number of objects and our tests have shown, moreover, that the robot is still capable of detecting the learned objects in its environment despite such difficulties.

One further point is that while our approach generalizes well even to other objects of the learned classes, our tests were inferior on popular datasets such as MS COCO.
This is due to the diversity of the images and the versatile situations depicted in them.
For instance, fruits such as bananas and apples can be found in their natural form as well as cut into small pieces in a fruit salad.
This task can only be solved with an enormous amount of training data.
Our method, on the other hand, although it cannot achieve the performance of training on large datasets, is primarily designed to teach the robot objects for which data does not yet exist.
In such scenarios, a semantic scene understanding is necessary and humans can assist in gaining this understanding by means of our method.
While an extension of existing datasets would also be conceivable, our method, in contrast, does not require tremendous labeling resources and can be used spontaneously in the respective situation.
Our evaluations (\mbox{\tblref{tbl:evalClasses}}) show that our method is capable of learning unknown objects that are not detected by the baseline.
It can therefore be used more flexibly without the need to know the situation in advance and rely on the existence of appropriate datasets.
Moreover, teaching through two-way interaction is extremely natural, especially since the AR system enables real-time communication between human and robot by directly connecting both worlds, the analog world of the human and the digital world of the robot, so that the human can supply information (gaze, class) to the robot, thus initiating the recording process, and the robot can in turn communicate feedback visually.
Pointing to objects using gaze completes the interplay, as it is intuitive, less ambiguous than pointing with a finger and, unlike speech, can be applied before the object is known to the robot.
All in all, we therefore consider our approach to be less of a replacement and more of a supplement.
	\section{Conclusion}
In this work, we presented a novel pipeline towards the deployment of robots in non-predefined scenarios.
To this end, we leveraged human gaze and augmented reality in the interaction between robot and human to successfully teach the robot new, yet unknown objects in its environment.

In order for robots equipped with machine learning based object detectors to detect their environment and the objects contained therein, a lot of training data is usually required.
In practice, however, under unpredictable conditions and due to the wide range of existing objects, the availability of a suitable data set can not always be guaranteed.
Our approach can complement popular datasets in exactly such situations and produce large amounts of automatically labeled, non-synthetic training data in a user-friendly manner and in a short period of time.
On the basis of such data, we have trained state-of-the-art object detectors in several different ways and shown that it is possible to learn and detect new objects in this manner.
In fact, the training can even take place standalone on the robot due to transfer learning, without the need for tremendous computational resources.
Further, with a few instances, it was possible for the robot to generalize to unseen objects in the given class and to detect classes that could not be detected by the baseline due to an unsuitable underlying training dataset.
This makes our teaching pipeline a valuable extension to training exclusively on standard datasets.
Overall, our approach is supremely natural and intuitive by virtue of its multimodality, including AR and the shared gaze of human and robot.
The dataset we have recorded in the course of our evaluation is also made publicly available and is characterized by a high level of information density owing to the many different perspectives on the respective object and the data gathered in this process.
As a result, it has the potential to be relevant for a variety of purposes, aside from ours.

However, a significant amount of work remains for the future as we plan to investigate the usability of our system in a user study as well as to extend our approach to enable the robot to successfully detect objects outside of its trained environment and to further leverage the acquired knowledge through active learning in another human-robot interaction scenario.

	\section*{Acknowledgments}
%Funded by the Deutsche Forschungsgemeinschaft (DFG, German Research Foundation) under Germany’s Excellence Strategy -- EXC number 2064/1 -- Project number 390727645.
This research was funded by the Deutsche Forschungsgemeinschaft (DFG, German Research Foundation) under Germany's Excellence Strategy -- EXC number 2064/1 -- Project number 390727645.
We also thank Julia Dietl for her valuable efforts in labeling.
	
	\bibliographystyle{plain}
	\bibliography{literature}
	
\end{document}